\documentclass[12pt]{article}

\usepackage{sbc-template}
\usepackage[T1]{fontenc}
\usepackage[utf8]{inputenc}
\usepackage{graphicx}
\usepackage{url}
\usepackage{natbib}
\usepackage{hyperref}
\usepackage{algorithm}

\newcommand{\mytitle}{Decentralized   Validation   for   Non-malicious
  Arbitrary Fault Tolerance in Paxos}

\hypersetup{
     pdftitle={\mytitle}, 
     pdfauthor={Rodrigo R. Barbieri, Enrique S. dos Santos, Gustavo M. D. Vieira},
     pdfdisplaydoctitle=true,
     hidelinks
}

\bibpunct{[}{]}{,}{a}{}{,}
\renewcommand{\cite}[2][]{\citep[#1]{#2}}

\title{\mytitle}

\author{Rodrigo R.  Barbieri\inst{2} \and  Enrique S.  dos Santos\inst{1}
  \and Gustavo M.  D.  Vieira\inst{1} }

\address{DComp -- CCGT -- UFSCar\\
         Sorocaba, São Paulo, Brasil
         \email{enrique.santos@dcomp.sor.ufscar.br,
           gdvieira@ufscar.br}
         \nextinstitute
         Flextronics Instituto de Tecnologia\\
          Sorocaba, São Paulo, Brasil
          \email{rodrigo.barbieri2010@gmail.com}}

\begin{document}

\maketitle

\begin{abstract}
  Fault-tolerant  distributed systems  offer high  reliability because
  even  if faults  in  their  components occur,  they  do not  exhibit
  erroneous behavior.  Depending on  the fault model adopted, hardware
  and software  errors that do  not result  in a process  crashing are
  usually not tolerated.  To tolerate these rather common failures the
  usual  solution is  to adopt  a stronger  fault model,  such as  the
  arbitrary  or Byzantine  fault model.   Algorithms created  for this
  fault model, however, are considerably more complex and require more
  system  resources than  the  ones developed  for  less strict  fault
  models.  One approach to reach  a middle ground is the non-malicious
  arbitrary fault model.  This model assumes it is  possible to detect
  and filter faults with a given  probability, if these faults are not
  created with malicious intent, allowing the isolation and mapping of
  these faults  to benign  faults. In  this paper  we describe  how we
  incremented  an   implementation  of   active  replication   in  the
  non-malicious  fault   model  with  a  basic   type  of  distributed
  validation, where  a deviation from the  expected algorithm behavior
  will  make  a  process   crash.   We  experimentally  evaluate  this
  implementation using a fault injection  framework showing that it is
  feasible  to extend  the  concept of  non-malicious failures  beyond
  hardware failures.
\end{abstract}

\section{Introduction}

Fault-tolerant distributed systems offer high reliability because even
if faults  in their  components occur, they  do not  exhibit erroneous
behavior. Depending on the fault  model adopted, common faults such as
message loss and  processes crashes have no effect  on the distributed
algorithm being run. However, hardware and software errors that do not
result in  a process crashing  are usually  not tolerated in  the most
used fault models.  Considering the hardware, examples of these faults
are memory corruption caused by spontaneous bit-flips, disk corruption
caused by media defects, message corruption due to interference, among
others. Looking  at the  software, programmer error,  operating system
bugs  and operator  error can  violate the  assumptions of  the target
fault model.

To  tolerate these  rather common  failures the  usual solution  is to
adopt  a stronger  fault  model that  assumes any  type  of fault  can
occur. The  \emph{arbitrary} or Byzantine  fault model~\cite{generals}
is very well  understood and in fact offers tolerance  to a wide range
of fault types.  Algorithms created for this fault model, however, are
considerably more complex  and require more system  resources than the
ones  developed  for less  strict  fault  models. For  instance,  many
arbitrary fault model  algorithms require that \emph{at  most} a third
of   the   components   of   the    system   present   any   type   of
fault~\cite{castroPracticalByzantine}.     This   is    a   reasonable
requirement    considering   the    strength   of    fault   tolerance
provided. Nonetheless, this implies that  to achieve the same level of
fault  tolerance a  system designer  would have  to use  a third  more
machines than necessary for basic crash tolerance. Worse still, if one
considers programming error as the  more likely fault to be tolerated,
no  more than  one third  of the  system is  allowed to  use the  same
implementation.  This happens because  if a single implementation were
used in the entire system,  a programming error in this implementation
would generate faulty behavior in more than a third of the machines.

The  choice of  fault  model can  be considered  a  trade off  between
stronger fault  tolerance and  simpler implementation  requiring fewer
resources. In a sense, the gap between crash-tolerant fault models and
arbitrary fault model is  too large~\cite{nonMalicious}.  One approach
to  reach  a  middle  ground,  focused  on  hardware  faults,  is  the
\emph{non-malicious  arbitrary} fault  model~\cite{nonMalicious}. This
model assumes it is possible to  detect and filter faults with a given
probability, if  these faults are  not created with  malicious intent,
allowing  the  isolation  and  mapping   of  these  faults  to  benign
faults. In  practice this means that  it is possible to  use a generic
adapter   that   allows    crash-tolerant   algorithms   to   tolerate
non-malicious arbitrary faults.

We  believe it  is  possible  to go  further  than detecting  hardware
faults,  and  detect other  faults  that  are non-malicious,  such  as
programming and configuration faults. By its non-malicious nature, the
effects of  these failures will  be probabilistic detectable,  and the
system can be made  tolerant to them.  To this end we  have as a first
step modified an implementation of active replication~\cite{schneider}
as  a testbed  for tolerating  failures in  a non-malicious  arbitrary
setting. We were successful in  hardening the testbed against hardware
failure and local data corruption~\cite{barbieri15}.

In  this  paper  we  describe  how we  implemented  a  basic  type  of
distributed validation, where a  deviation from the expected algorithm
behavior will make  a process crash.  We  also experimentally evaluate
the hardened testbed  using a fault injection  framework.  Our results
show  that it  is  feasible  to extend  the  concept of  non-malicious
failures  beyond hardware  failures.  The  main contributions  of this
paper  are the  description of  the distributed  validation used,  the
protocol to validate its effectiveness and evidence that this approach
is  sound.    The  paper  starts   by  defining  more   precisely  the
non-malicious arbitrary fault model in Section~\ref{sec:model} and how
to implement algorithms for it in Section~\ref{sec:nonmalicius}. Then,
we describe  our testbed  and our  proposed distributed  validation in
Section~\ref{sec:hardtreplica}.    We   close   the  paper   with   an
experimental     evaluation      of     our      implementation     in
Section~\ref{sec:experiment}.


\section{Fault Model}
\label{sec:model}

Fault models abstract  the properties a system must satisfy  and how a
distributed algorithm  should tolerate  faults. Two very  common fault
models  in   which  distributed   algorithms  are  designed   are  the
\emph{crash-stop}  and \emph{crash-recovery}  models.  In  both models
processes only fail by completely  crashing.  We can call these models
\emph{benign} because it  is assumed that the  tolerated failures will
respect  a (probably  unknown) probability  distribution.  A  stronger
fault   model   that  assumes   more   types   of  failures   is   the
\emph{Byzantine}  or \emph{arbitrary}  model  in  which processes  can
deviate in any way from the algorithm specification.

In the arbitrary fault model it  is impossible for processes to decide
whether  another  process  is behaving  arbitrarily  intentionally  or
not. This  assumption covers  virtually any type  of failure  a system
might encounter. We refer to \emph{malicious} faults when a process is
behaving  arbitrarily  intentionally,   through  manipulation  from  a
malicious  agent. These  type of  faults do  not follow  a pre-defined
probability distribution and  can occur in response  to measures taken
to tolerate them.

\begin{figure}[htbp]
\centering
\includegraphics[width=0.4\textwidth]{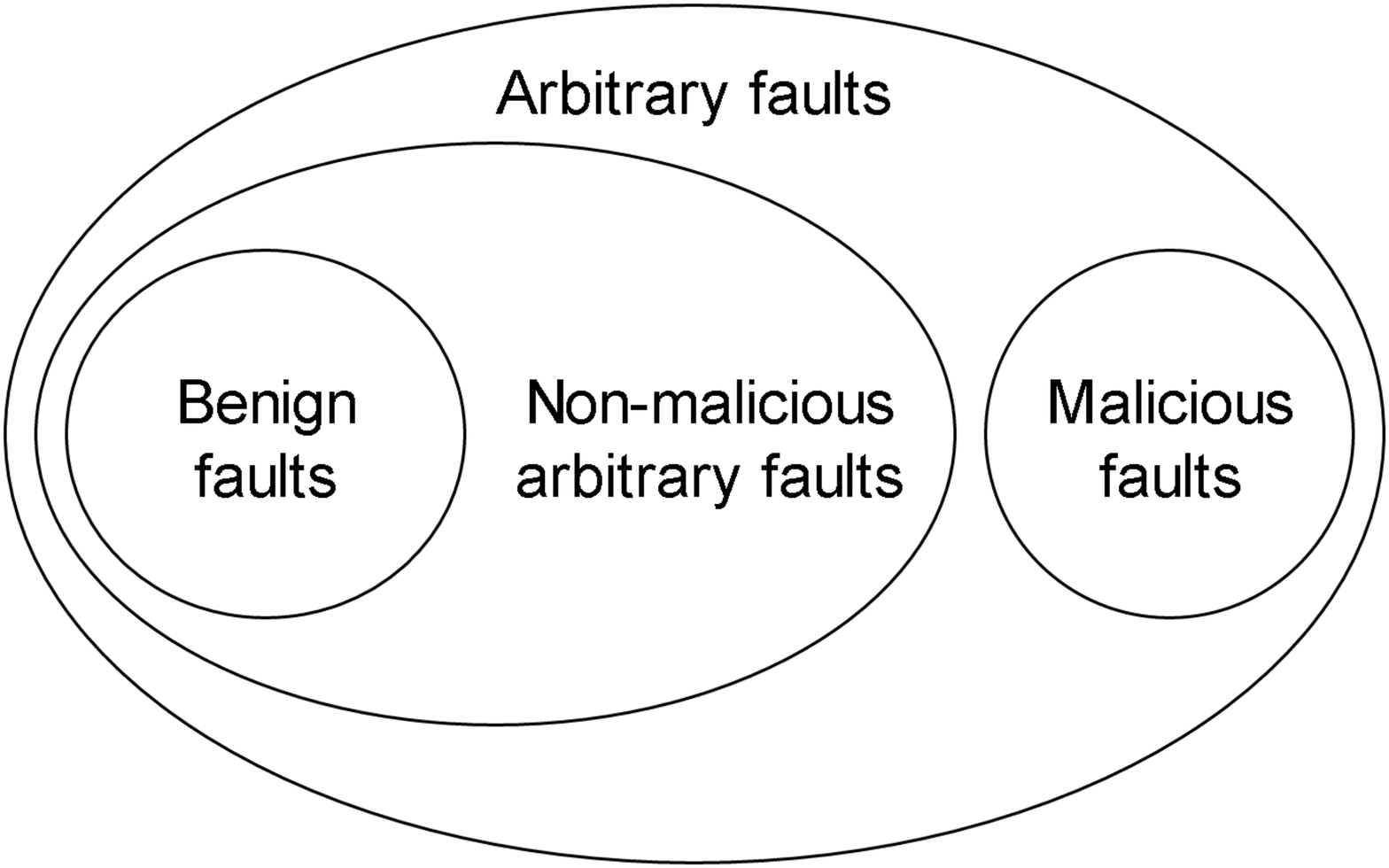}
\caption{Relationship among fault classes}
\label{fig:models}
\end{figure}

Fault  models  range  from  weaker (more  strict)  to  stronger  (more
general),  as  shown  in Figure~\ref{fig:models}.   The  stronger  the
model,  the  more  complex  and   difficult  it  is  to  implement  an
algorithm.  When  building  a  practical  distributed  system,  it  is
desirable  to adopt  a fault  model that  better fits  the system  and
satisfies its requirements for performance and types of faults it must
tolerate.  However, this is not always the case, since any distributed
system that relies on actual computers is prone to arbitrary faults.

Distributed system designers desire  to tolerate arbitrary faults, but
would prefer a  less performance intensive algorithm  than a byzantine
one~\cite{datacenterNonMalicious,  hardening,   nonMalicious}.   While
malicious    faults    are    being    tolerated    using    different
techniques~\cite{datacenterNonMalicious, hardening}, and  based on the
premise  that  any  fault  model  can be  hardened  to  tolerate  some
arbitrary faults, it  is possible to harden  the crash-recovery benign
model  to tolerate  non-malicious arbitrary  faults, thus  achieving a
fault model similar to the  arbitrary fault model, but where malicious
faults are not necessarily tolerated by the algorithm.

An  algorithm  for the  non-malicious  arbitrary  fault model  can  be
considered to be less complex  than an classical arbitrary one because
it does not tolerate malicious faults in its implementation.  However,
the implementation  required to  tolerate all  non-malicious arbitrary
faults adds its own complexity to  the algorithm.  This fault model is
more   precisely   described   in   terms   of   the   two   following
properties~\cite{nonMalicious}:

\begin{description}
\item  [No   impersonation:]  the  environment  never   creates  valid
  messages,  except for  duplicates. This  property assumes  that only
  processes themselves are able to create valid messages, so malicious
  agents cannot interact with existing processes in a system.
\item [No propagation:] a process that is considered faulty, by either
  itself or  by another process,  cannot ever create a  valid message.
  This property assumes that when the process has become faulty, it is
  not allowed  to send any more  messages, nor any correct  process is
  allowed to accept messages from a faulty process.
\end{description}

An  algorithm for  this fault  model  is expected  to tolerate  faults
caused by data corruption, such as from persistent memory, main memory
or network, in addition to benign faults~\cite{datacenterNonMalicious,
  hardening, nonMalicious}.

\section{Tolerating Non-malicious Arbitrary Faults}
\label{sec:nonmalicius}

Non-malicious  arbitrary  fault types  are  present  not only  in  any
practical  distributed  system,  but  in any  system  that  relies  on
computer  components that  can fail.   These faults  can be  tolerated
through  error  detection techniques,  such  as  integrity checks  and
semantic checks.   The techniques described  below aim to  detect data
corruption,  memory  corruption,  and  in some  cases  programmer  and
operator mistakes.  However, each  approach mentioned has its overhead
cost   associated,  for   either   performing   repeated  checks   and
recalculations,  encrypting, or  doubling memory  requirements due  to
redundancy.

\subsection{Integrity checks}

Integrity checks  verify data  by saving at  least one  redundant copy
that  can be  used  to validate  against the  original  data, such  as
checksums,  duplicate states,  timestamps or  data size  values.  This
approach  is commonly  used when  reading and  writing data  from main
memory, storage  and peer-to-peer  network message exchanges.   We now
describe in more detail these techniques:

\begin{description}
\item [Data  and state  redundancy]: each  process variable  or stored
  data has  a duplicate which  can be  validated against and  used for
  backup. The  duplicates must be always  be kept in sync  and checked
  for consistency  on each  read and  write operation.   This approach
  clearly uses  a significant amount  of additional memory and  has an
  increased     overhead     for     keeping    both     states     in
  sync~\cite{datacenterNonMalicious, hardening, paxosMadeLive};
\item  [Checksums  and  hashes   redundancy]:  the  usage  of  encoded
  redundancy allows for future detection of undesired corruption.  The
  most common type  of redundancy is generating a checksum  or hash of
  data and attaching it to the protocol messages prior to transmitting
  across the  network or saving  them in  storage.  Any read  or write
  operation on data  must recalculate the checksum  and verify against
  the one  attached to the  message, adding a  significant performance
  cost       related       to       the       checksum       algorithm
  used~\cite{datacenterNonMalicious, hardening, paxosMadeLive};
\item [Encoding and arithmetic codes]:  the usage of in-place encoding
  and  decoding, like  numerical properties  of data,  can be  used to
  detect undesired  corruption in each  read and write  operation. For
  instance, if  numerical variables are  multiplied by a  prime number
  upon writing and divided by the same number when they are read back,
  the remainder should always be zero.  This approach is considered to
  be      very     efficient      performance-wise,     but      lacks
  coverage~\cite{nonMalicious}.
\end{description}

\subsection{Semantic checks}

Semantic checks validate  that after an operation has  been applied on
data, the  newly obtained state  is semantically correct  according to
the  applied  operation~\cite{datacenterNonMalicious}.  For  instance:
after adding  an element  to a list,  check if the  element is  in the
list.  This approach  has  the  added benefit  of  testing the  system
against possible bugs, which was  one scenario in the experiment found
in~\cite{paxosMadeLive}.

\section{Hardened Treplica}
\label{sec:hardtreplica}

\subsection{Treplica and Paxos}
\label{sec:paxos}

Treplica~\cite{vieira08a,vieira-tr10b} is a Java framework that allows
distributed applications  to use Paxos  as middleware to  manage state
replication through its state machine.  Its implementation is close to
the   Multi-Paxos  approach~\cite{lamport}   with  a   few  additional
optimizations, like Fast  Paxos support~\cite{fastPaxos} and broadcast
votes,  where each  learner agent  receiving a  majority of  votes can
commit the change immediately.

\begin{figure}[htbp]
\centering
\includegraphics[width=0.8\textwidth]{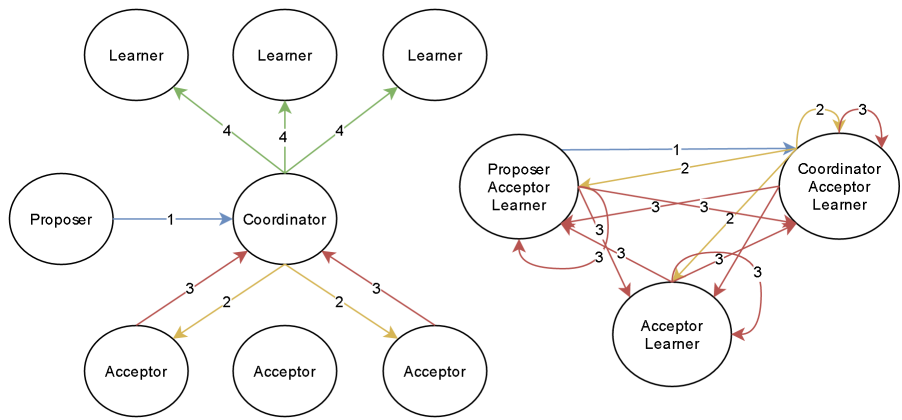}
\caption{Multi-Paxos (left) and Multi-Paxos in Treplica (right)}
\label{fig:treplica_paxos}
\end{figure}

In Treplica, replicas can concurrently perform any Paxos role, such as
coordinator, proposer, learner and acceptor.  This is analogue to many
practical  middlewares  implementing  Paxos, and  allows  for  greater
flexibility    in    the    amount     of    replicas    and    system
configurations.       Figure~\ref{fig:treplica_paxos}      illustrates
Multi-Paxos algorithm  messages exchanged during  a common round  in a
consensus instance,  highlighting the differences  between theoretical
Multi-Paxos and  Treplica's implementation.  The message roles  are as
follows:

\begin{description}
\item[Message \#1:] Client proposal message sent to coordinator;
\item[Message \#2:] Proposal sent to acceptors for voting;
\item[Message \#3:] Acceptors vote on proposal;
\item[Message \#4:] Decision is broadcast to learners.
\end{description}

In Treplica, voting messages, labeled  \#3 in the figure, are received
by learners and the proposer as  well, thus allowing learners to apply
the state  transition immediately. Also,  the proposer can  respond to
the client as soon as receiving a majority of votes. Moreover, message
\#4 is not necessary  but is used to broadcast a  decision if there is
any message loss.

Applications designed according  to the Model-View-Controller standard
can easily be  modeled to use Treplica. We chose  Treplica because its
modular architecture  allows for improvements  to be easily  coded and
tested. Since it is designed  to tolerate benign faults, upon analysis
we validated that it is prone to non-malicious arbitrary faults we are
interested in.   Additionally, Treplica  is object-oriented  and makes
use  of immutable  objects design,  where an  object is  never changed
after  being  instantiated. This  allows  for  more efficient  use  of
checksums. State transition  semantic checks can also  be easily coded
by  the application  due to  its  integration with  the state  machine
modeling.

\subsection{Tolerating non-malicious arbitrary faults}

Our main approach to harden the benign crash-recovery fault model into
the  non-malicious   arbitrary  one  is  to   employ  fault  detection
techniques, as  described in Section~\ref{sec:nonmalicius},  to detect
faults resulting  from data  corruption, while initially  not worrying
about  how  to  recover  from  them.   Moreover,  we  have  created  a
distributed validation mechanism to try and detect deviations from the
expected algorithm properties, that is, if the replicas state start to
diverge.  Upon detecting  unexpected faults, our proposal  is to abort
the replica execution, preventing it from propagating corrupt data and
further  participating  in  the consensus  algorithm.   This  strategy
reduces the  crash-recovery fault model  to a simpler  crash-stop one,
but tolerating non-malicious arbitrary faults.  This is similar to the
approach  taken by~\cite{nonMalicious},  but adds  the ability  to
detect protocol violations.

From the  point of view  of a  benign fault model  distributed system,
most arbitrary  faults behave  as silent  faults because  their errors
cannot be easily detected. For instance,  if a user clicks a button to
buy one book, but a replica  processes that two books have been bought
because bits got flipped along the way, then this is not an error from
Paxos point  of view, because  the message was  delivered consistently
across all  replicas.  In  order to  effectively tolerate  such silent
faults, we employed the following techniques:

\begin{itemize}
\item Integrity checks, to address data corruption.
\item State checks, to address state corruption.
\item Semantic checks, to address programmer mistakes in the application.
\item Distributed  validation, to address data  corruption, programmer
  or configuration faults.
\end{itemize}

Integrity, state  and semantic  checks were  implemented based  on the
ideas  described  in  Section~\ref{sec:nonmalicius}. Details  of  this
implementation   can   be    found   in~\cite{barbieri15},   and   are
unfortunately omitted from this paper because of lack of space.  Next,
we  describe in  detail  our proposal  for  distributed validation  of
replication algorithms.

\subsection{Distributed validation}
\label{sec:distributed}

In  a distributed  system in  which each  replica has  its independent
state, although Paxos  can guarantee state transitions  are applied in
the same  order, it cannot guarantee  that all replicas will  have the
same  state in  the  presence of  non-malicious  arbitrary faults.   A
replica  that  experiences  an  arbitrary fault  may  have  its  state
diverged from the others, while state transitions will continuously be
applied on  top of  the corrupt  state. This may  allow the  system to
display  incorrect  data when  the  replica  is  queried by  a  client
application.  This divergence can happen by a corruption in the memory
holding the \emph{code} of the protocol  or, more likely, be caused by
programming or  configuration fault.   In order to  completely satisfy
the ``No propagation'' property of the fault model, this scenario must
be covered by a validation mechanism.

A distributed  validation mechanism allows replicas  to validate their
state upon receiving  a network message containing a  checksum or hash
that  is related  to  the  state they  currently  are, thus  detecting
possible state  divergences. A state  of a replica comprises  both the
application  state and  the internal  state of  all Paxos  agents.  We
developed a way to use the Paxos algorithm to perform this validation,
thus having the algorithm extended with this mechanism.  We attempt to
detect  state  divergence  between  replicas by  including  the  state
checksum  in the  \emph{voting}  messages exchanged  by  Paxos in  the
accepting phase (Message \#3, see Section~\ref{sec:paxos}).  Acceptors
read the  checksum from  the application  when creating  the immutable
voting  messages and  attach it  to them.   In Treplica,  all replicas
receive the  voting messages,  thus the  learner module  validates the
local state upon receiving them using the attached checksum.

In order to minimize the performance impact and adapt the mechanism to
Treplica's  architecture,   we  decided   to  take  an   eventual  and
opportunistic  validation  approach. By  defining  a  window of  state
counters in which the state checksum is updated, the replicas are able
to  eventually  validate  a  state within  the  defined  window.   For
instance, if  the window value  is 100,  then the state  checksum will
change   only  every   one   hundred  state   transitions  have   been
applied. This makes it easier to  synchronize all replicas in the same
window, because Paxos, and many other distributed algorithms, does not
impose a hard limit on the  speed differences of the system processes.
The only requirement is that a quorum be approximately synchronized, a
single  straggler process  can be  in a  state arbitrarily  behind the
others.

\begin{figure}[htbp]
  \centering
  \includegraphics[width=0.8\textwidth]{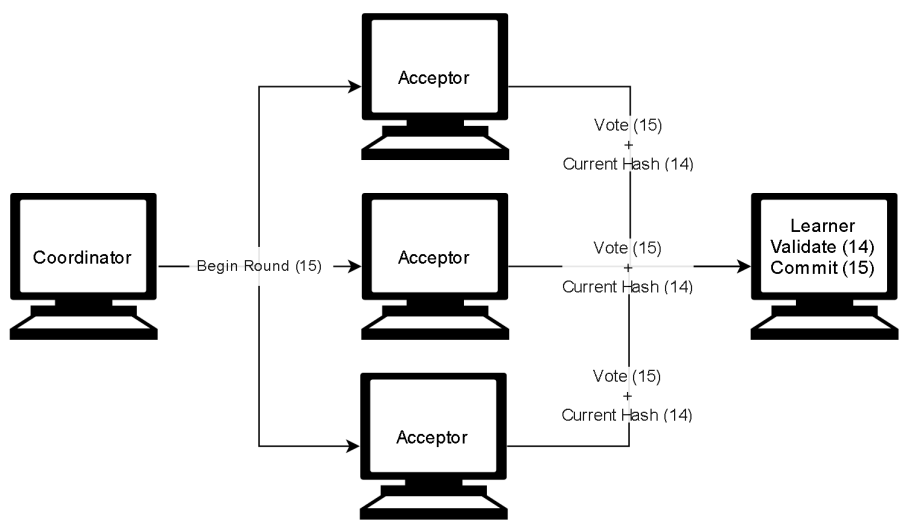}
  \caption{Distributed validation mechanism implemented in Treplica}
  \label{fig:distributed_validation}
\end{figure}

In  Figure~\ref{fig:distributed_validation},  acceptors include  their
state  checksum numbered  \#14 when  consensus instance  of the  state
transition numbered \#15 is running. Both consensus instances \#14 and
\#15 are  related to the same  window, which is from  state transition
\#1 to state transition \#100,  thus they carry the checksum generated
in transition \#1.  The learner validates state checksum numbered \#14
before committing the state transition numbered \#15. In this example,
if any of the replicas have  their state diverging within this window,
it  will  be detectable  only  after  transition  \#100, where  a  new
checksum will be included in  exchanged messages.  Without this window
mechanism, replicas  would rarely be  able to validate  received state
checksums  related to  the same  state count,  because they  apply the
state  transitions in  an asynchronous  way. Treplica  packs a  set of
state  transitions in  the same  Paxos instance,  and replicas  end up
advancing rounds in  different paces, resulting in  the current window
and backlog  variables frequently getting  discarded due to  the state
count advancing before having a chance to validate.

We  defined   two  variables  where   we  store  the   received  state
checksums. One  that is related to  the current window, and  a backlog
one that  is related  to the  next window to  be processed.   The next
window is determined  by the first received message that  does not fit
into the current  window. Messages received that fit  into the current
window  are  validated  immediately,  while messages  related  to  the
subsequent registered window  are stored for later  validation. When a
replica advances to the window  that contains messages to be processed
later,  it  moves  all  the  stored messages  to  the  current  window
variable, clears the backlog one and starts processing them.

Validation  consists  in  comparing  the  validating  replica's  state
checksum to the  majority of received state checksums. As  soon as the
number of received common state checksums is the same as the number of
replicas  in  the  majority  quorum  in  the  system,  the  validating
replica's state checksum is compared to this common state checksum. If
the  validating  replica's state  checksum  is  not  the same  as  the
majority, then the validating replica detects that it has diverged and
aborts execution.

In  Algorithm~\ref{algorithm:distributed_validation} it  is shown  the
validation code  that is performed  by learners. Every  state checksum
received that is not related to  the current window is saved for later
processing. When a  state checksum that matches the  current window is
received, it  is saved  in a structure  responsible for  storing state
checksums  indexed by  replica  unique  identification numbers.   This
structure  has   a  method  getMostCommonChecksum()   responsible  for
returning  a  list of  the  most  common  occurrence for  the  current
window. If  the size of  this list matches  the quorum size,  then the
learner validates its own checksum against that common checksum, which
raises an exception if it diverges.

\begin{algorithm}
\caption{Distributed validation}\label{algorithm:distributed_validation}
\footnotesize
\begin{verbatim}
void receiveVotingMessage(Message message){
    StateChecksum checksum = message.stateChecksum;
    if (checksum.stateCount == Application.currentChecksum.stateCount){
        saveAndProcessStateChecksum(checksum);
    } else {
        if (backlog.stateCount == checksum.stateCount){
            saveMessageInBacklog(checksum);
        }
    }
    processVotingMessage(message);
    
}

void saveAndProcessStateChecksum(StateChecksum checksum){
    currentWindow.saveMessageInCurrentWindow(checksum);
    List<StateChecksum> list = currentWindow.getMostCommonChecksum();
    if (list.lengh == Application.quorumSize){
        if (checksum.hash != list.get(0).hash){
            throw new StateDivergedException(
                    Application.currentChecksum);
        }
    }
}

\end{verbatim}
\end{algorithm}

Expecting  a  majority is  a  way  to  avoid  a correct  process  from
terminating as it encounters the state  from a corrupted process. If a
majority of  processes has the same  state checksum and this  value is
different from the local state checksum,  the process is certain to be
one that diverged.   We consider this mechanism to  be eventual, since
replicas may not participate in  certain voting rounds. Moreover, this
approach would fail in case a majority of replicas diverge in the same
window.  Such  mechanism would still  depend on the  application being
able to generate  a checksum of its  state or of certain  data that is
comparable to other  replicas. The more precise  this information, the
more coverage this mechanism can achieve.

\section{Experimental Validation}
\label{sec:experiment}

To  evaluate  the  dependability  of hardened  Treplica,  we  injected
randomly generated faults in a test system and analyzed the occurrence
of errors.   Our testing  system consists  in a set  of replicas  of a
distributed  application, implemented  on  top of  Treplica.  We  have
generated  load  to  simulate  user requests  to  each  replica  while
injecting  failures, and  we  have  measured the  error  ratio of  the
hardened  Treplica.   In this  section  we  describe our  experimental
setup,  how  faults  were  injected, the  parameters  of  these  fault
injections, and the observed error rate.

\subsection{Test Setup}

In order  to generate  the load,  the test  system running  an example
application received  requests which cause state  transitions for five
minutes.  We created a manager application  to run 50 instances of the
same test, preparing and cleaning each replica instance resources used
by each  instance, such as storage  folders and logs. When  starting a
test instance the  manager application starts the  replicas and starts
the load generating application, which  sends the requests. In case an
error  is detected,  the application  crashes or  times out,  then the
manager  application aborts  the  test execution,  cleans the  storage
folders and logs the results to  a separate folder indexed by the test
instance count.  Our request timeout was set to five minutes, in order
to  detect  whether a  failure  made  the  system unable  to  continue
processing requests.

We coded  a simple  application built  on top  of Treplica  to perform
tests: a hash  table of strings.  Client requests can  add, remove and
list   elements    in   the   hash.    In    this   application,   the
application-defined state information implemented counts the number of
elements  in  the  hash.   The semantic  checks  implemented  for  the
``AddElement'' operation validates if the  added element is present in
the hash,  while for the  ``RemoveElement'' operation it  validates if
the element  is not present in  the hash.  Each request  was always an
``AddElement'' operation,  where the element consisted  of a Long-type
counter  value converted  to  string value,  appended  by the  replica
identification number.  When running tests with fault injections using
this application, we always printed all  the elements in the hash when
the test finished without detected faults to get a record of the final
state of the replica.

Instead  of running  Treplica on  separate machines  in a  cluster, we
opted to run  five instances of the same application  at the same time
in  a  single  machine.   Each  instance had  a  separate  folder  for
individual storage,  thus the  replicas were independent.   Running in
the same machine makes it easier  to overload the system, dropping and
reordering  messages, creating  configurations not  easily found  in a
system with  low load.  Curiously,  we actually  did run the  tests in
independent machines in a cluster to parallelize individual executions
and cut total time required for  the 50 runs of each experiment.  Each
node  in  the cluster  had  the  following configuration:  Intel  Xeon
E5-2665 2.40 GHz processor,  16 GB DDR3 1333 MHz RAM,  1 TB HDD, Linux
CentOS 6,  Java Development Kit  (JDK) version 1.8.0 update  191.  The
specific  amount of  requests per  second we  chose was  calibrated to
overload  it,  creating  messages   reordering  and  loss.   For  this
particular  hardware we  settled with  10000 Op/s  per replica,  for a
total of 50000 Op/s total.

\subsection{Fault Injection using AspectJ}

To inject faults we used  an aspect-oriented library known as AspectJ.
This library  allows us  to change  the behavior  of any  Java program
without        changing        its        main        code.         In
Algorithm~\ref{algorithm:example_injection}  it  is shown  an  example
injection, where in order to inject a fault in the operation of adding
a string  to a  list, a method  must be created  to have  its behavior
overridden. The method in this case is ``listAdd''. Our injection code
runs instead of  the original code every time  ``listAdd'' is invoked,
we then use a local variable to  decide whether we inject a fault that
consists in  running the original  function with a  different argument
value, or  if we allow  the function  to continue without  faults. The
idea is to ``corrupt'' the execution of the method as if its arguments
were subject to bit flips or memory corruption.

\begin{algorithm}
\caption{Example injection}\label{algorithm:example_injection}
\footnotesize
\begin{verbatim}
private List<String> list;
boolean inject;

boolean listAdd(String name){
    list.add(name);
}

boolean around(String name) : listAdd(name){
    if (inject)
        return proceed(random.nextLong().toString());
    else
        return proceed(name);
}
\end{verbatim}
\end{algorithm}

From  this basic  idea, we  created a  fault injection  framework that
reads a configuration  file when Treplica is initialized  and sets the
fault injection conditions and modes according to a set of parameters.
Whenever an aspect-marked method in the original code is executed, the
aspect  code is  invoked.  If  the condition  is not  met, no  code is
injected and  the original code  is run.   The two modes  of operation
implemented are:

\begin{description}
\item [Single  timed injection mode:] in  this mode, a timer  value is
  defined in the configuration file for each injection. Once the timer
  expires, the condition allows the fault to be injected. The fault is
  injected the  next time the  aspect is  invoked.  Once the  fault is
  injected, the condition is permanently disabled. This mode is useful
  when we want to confirm that a single fault was tolerated or not;
\item [Probability-based injection mode:]  in this mode, a probability
  value   is   defined   in    the   configuration   file   for   each
  injection. Whenever the aspect-marked  code is executed, it randomly
  generates a probability value. If  it is higher than the probability
  value defined, then the condition for fault injection is met and the
  fault  is injected  immediately.  All  subsequent executions  of the
  aspect-marked code will check  against the probability of injection,
  which may inject more fault occurrences. This mode is useful when we
  cannot guarantee that a single fault injection can cause an error to
  occur, because  it may  require specific conditions  to be  met. For
  example, a failure may require many consecutive messages to be lost,
  else the errors will be absorbed by the algorithm.
\end{description}

\subsubsection{Injection Scenarios}

With the framework  in place, we choose three sets  of specific faults
to  inject in  the replicas:  message faults,  application faults  and
algorithm  faults.   Message  injections  corrupts  data  arriving  or
leaving the  address space  of the replica,  such as  network messages
exchanged between replicas, log of operations saved and retrieved from
secondary  storage, and  checkpoints,  also saved  and retrieved  from
secondary storage. Application injections  corrupts the internal state
of the  application. Algorithm  injections makes the  algorithm behave
differently from its specification.

For message injections, we used single timed injection mode to corrupt
messages in transit.  In network,  storage and checkpoint messages, we
change a random  value in the message  as soon as it  is received from
the network  or recovered from  storage, while retaining  the checksum
value on the  hardened Treplica.  For application  injections, we used
single timed injection  mode to inject faults  during state transition
in the hash of strings application.   We inject faults in the state by
manipulating  the state  transition operation  to either  not add  the
elements or  by changing  their string values.   For both  message and
application injections  the fault  was injected  10 seconds  after the
test started.
    
For  algorithm injections,  we used  probability-based injection  mode
attempting to  cause failures  by breaking  simple but  very important
algorithm   mechanics,   in  order   to   cause   replica  states   to
diverge. Assuming the Paxos  algorithm is keeping replica consistency,
the faults injected are as follows:

\begin{description}
\item  [Learner commits  with no  quorum:] This  injection causes  the
  affected  learner to  have a  probability of  committing a  proposal
  without requiring a  quorum to do so. If the  round does not succeed
  and a  new value  is proposed,  the replica ends  up with  its state
  diverging from the others.
\item [Acceptor forgets its previous votes:] This injection causes the
  affected acceptor to  have a probability of  returning no previously
  registered votes for a given round.  The coordinator, upon receiving
  this message, will assume that  no previous value has been proposed,
  and will  propose a new  one instead,  while the previous  value may
  have been decided and committed by learners.
\item [Coordinator  forgets last  proposals received:]  This injection
  causes  the coordinator  to  have a  probability  of forgetting  the
  previous  proposals he  receives  from acceptors,  thus choosing  to
  start a new round with a new value while the previous value may have
  been decided and committed by learners.
\end{description}

We used 80\%  fault injection probability for  the injection scenarios
being tested. The Paxos algorithm is surprisingly resilient, even when
the  injected faults  made it  behave as  if a  programmer had  made a
mistake, the algorithm gives the correct  answer most of the time. The
high probability of  fault injection coupled with the  overload of the
system are  required to  create the scenarios  in which  the algorithm
exhibit  an  error.  With  the  selected  parameters,  we had  a  high
probability at  least one replica  to commit an undecided  proposal in
each test run, causing a state divergence.

However, to assess the behavior  of hardened Treplica with more faults
we also ran tests where we performed Paxos injections in all replicas.
Those test runs  are labeled “All replicas” in the  table, whereas the
“Single  replica” labels  refer to  injections performed  in a  single
replica. We noticed  that the test runs where  we performed injections
in all replicas were more likely  to result in all replicas diverging.
Additionally,  we  analyzed  the  possibility of  the  fault  injected
causing failures  only to  the replica  which it  was injected,  or if
errors propagated.  In all our test  runs, at least one  execution had
error propagation.

\subsection{Injection Test Results}

Table~\ref{tab:hardened_treplica}  lists the  error rates  for message
and  application fault  injection tests  in hardened  Treplica.  These
faults model  hardware faults  that happen locally  in a  replica. For
each run, a  fault was injected and it was  detected by the integrity,
state or  semantic checks.   We confirmed the  detection of  all fault
occurrences  either   through  logging  or  replicas   aborting  their
executions upon detection.

\begin{table}
\centering
\caption{Message and application injection tests}
\label{tab:hardened_treplica}
\scalebox{0.775}{
\begin{tabular}{|l|p{0.8cm}|p{1.6cm}|p{1.6cm}|p{1cm}|p{1cm}|}
\hline
Test & Runs & Fault \newline injections & Fault \newline detections & Errors & Rate \\ \hline
Message injections - Fault on a protocol message received & 100 & 100 & 100 & 0 & 100\% \\ \hline
Application injections - Fault when adding element & 100 & 100 & 100 & 0 & 100\% \\ \hline
\end{tabular}
}
\end{table}

For   algorithm    injections,   the   error   rate    is   shown   in
Table~\ref{tab:paxos_injections}.  For the test runs where faults were
not detected, there were cases where  there were no errors, thus there
was no  state divergence  between any of  the replicas.   We confirmed
such scenario by comparing all the elements printed by each replica at
the end of  a test run. The  runs where there were  errors, however, a
majority  of  replicas diverged  to  different  states, rendering  the
distributed validation  incapable of detecting any  divergence for the
rest  of  the test  run.   These  cases  could  also be  confirmed  by
comparing the printed  elements.  We consider these  occurrences to be
beyond our technique's coverage capability at this moment.

\begin{table}
\centering
\caption{Algorithm injection tests}
\label{tab:paxos_injections}
\scalebox{0.8}{
\begin{tabular}{|p{7.3cm}|p{2cm}|p{1.8cm}|p{1.8cm}|p{1.4cm}|p{1cm}|}
\hline
Test & Runs & Fault \newline injections & Fault \newline detections & Errors & Rate \\ \hline
Learner commits with no quorum \newline Single replica & 50 & 23500 & 45 & 1 & 98\% \\ \hline
Acceptor forgets previous votes \newline Single replica & 50 & 3662900 & 0 & 0 & 100\% \\ \hline
Learner commits with no quorum \newline All replicas & 50 & 194500 & 23 & 3 & 94\% \\ \hline
Acceptor forgets previous votes \newline All replicas & 50 & 14584980 & 16 & 0 & 100\% \\ \hline
Coordinator forgets received proposals \newline All replicas & 50 &  15000 & 47 & 2 & 96\% \\ 
\hline
\end{tabular}
}
\end{table}

Hardened Treplica was able to detect faults in all the fault injection
scenarios, except  for when all  replicas got their state  diverged in
algorithm injection  scenarios.  The only scenario  we injected faults
and  we could  not  detect the  resulting  errors was  the  case of  a
majority  of replicas  (five replicas  in our  test suite),  diverging
while  in the  same state  count  window.  We  consider the  following
conditions are needed for this to happen:
\begin{itemize}
\item There must be several  consecutive messages lost while the fault
  is injected.   We validated that  without message loss,  there could
  not be any failure.
\item The majority of replicas needs  to diverge within the same state
  validation   window    (see   Section~\ref{sec:distributed}).    The
  divergence of  individual replicas can  be easily detected  once the
  validation window changes  and updates the state  checksum, if there
  is still a majority of correct replicas.
\end{itemize}

We   summarize    our   test    suite   results   and    analysis   in
Table~\ref{tab:coverage}, where  we check for each  injection scenario
we tested:
\begin{itemize}
\item If  a potential error  from the fault injected  was successfully
  detected and which conditions were necessary for this detection;
\item To which  fault model the fault occurrence can  be mapped to and
  what other replicas observe of the faulty replica;
\item If the error can be propagated and disrupt other replicas;
\item Our coverage rate for the given fault.
\end{itemize}

\begin{table}
\centering
\caption{Non-malicious fault class coverage}
\label{tab:coverage}
\scalebox{0.8}{
\begin{tabular}{|p{4cm}|p{3.3cm}|p{4.2cm}|p{3.7cm}|p{1cm}|}
\hline
Fault & Detected? & Fault model mapped to & Can be propagated? & Rate \\ \hline
Message injections (network) & Yes & Benign crash-recovery \newline (single message loss) & No & 100\% \\ \hline
Message injections (stable storage) & Yes & Benign crash-stop \newline  (replica unavailable) & No & 100\% \\ \hline
Application injections (memory corruption) & Yes, but requires application support & Benign crash-recovery \newline (replica restarted) & Not applicable & 100\% \\ \hline
Application injections (bugs) & Yes, but requires application support & Benign crash-stop \newline  (replica unavailable) & Not applicable & 100\% \\ \hline
Paxos injections & Mostly & Benign crash-stop \newline (replica unavailable) & Yes & 97,6\% \\
\hline
\end{tabular}
}
\end{table}

Upon detecting a failure, we abort the replica execution.  Our current
solution results  in a crash-stop non-malicious  arbitrary fault model
instead  of  an  ideal crash-recovery  non-malicious  arbitrary  fault
model.  We consider the non-malicious arbitrary crash-stop fault model
to  be   more  resilient  than  the   original  benign  crash-recovery
implementation.  If a  benign  fault  occurs, the  system  is able  to
recover  itself and  continue, but  if an  error from  a non-malicious
arbitrary  fault is  detected  and is  non-recoverable,  we abort  the
replica execution, preventing any propagation of erroneous behavior.

\section{Conclusion}

In this paper we described the non-malicious arbitrary fault model and
how we adapted  the implementation of active replication  found in the
Treplica  framework  to  use  it.  We  proposed  a  novel  distributed
validation mechanism that expands the scope of non-malicious arbitrary
faults tolerated.  Our  experimental evaluation has shown  a very good
coverage of protocol  deviation faults, reaching 97,6\% in  the sum of
all tests.

These results are  particularly good, because here we  report our very
first attempt to detect deviations from the distributed algorithm. The
idea is very simple and is prone  to fail to detect multiple faults if
they happen too  fast. We believe this is a  promising approach and we
intend to create a more robust distributed validation mechanism.

\section*{Acknowledgments}

This    research   was    partially    funded    by   the    Brazilian
\textsl{Coordenação de  Aperfeiçoamento de  Pessoal de  Nível Superior
  (CAPES)}   under  the   \textsl{Pró-Equipamentos}  program   (Edital
25/2011).

\bibliographystyle{apalike}
\bibliography{non.malicius.crash}

\end{document}